\documentclass[aps,prl,twocolumn,groupedaddress,preprintnumbers,
showpacs]{revtex4}
\usepackage{graphicx}% Include figure files

\begin{document}
\preprint{APS/123-QED}

\title{Dynamic Point-Formation in Dielectric Fluids}
\author{Cheng Yang}
\affiliation{The James Franck Institute and Department of Physics, The
  University of Chicago, 5640 S. Ellis Ave., Chicago, IL 60637} 
\date{\today}
\begin{abstract}
We use boundary-integral methods to compute the time-dependent
deformation of a drop of dielectric fluid immersed in another dielectric
fluid in a uniform electric field $E$. Steady state theory predicts,
when the permittivity ratio, $\beta$, is large enough, a  
conical interface can exist at two cone angles,
with $\theta_<(\beta)$ stable and $\theta_>(\beta)$ unstable. Our
numerical evidence instead shows a dynamical process which produces a
cone-formation and a transient finite-time singularity, when $E$ and
$\beta$ are above their critical values. Based on a scaling analysis
of the electric stress and the fluid motion, we are able to apply 
approximate boundary conditions to compute the evolution of the tip
region. We find in our non-equilibrium case where
the electric stress is substantially larger than the surface tension, the
ratio of the electric stress to the surface tension in the newly-grown
cone region can converge to a $\beta$ dependent value,
$\alpha_c(\beta)>1$, while the cone angle converges to
$\theta_<(\beta)$. This new dynamical solution is self-similar. 
\end{abstract}

% Computational methods in fluid dynamics, hydrodynamic stability,
% liquid-liquid interface
\maketitle
The formation of conical ends on fluid-fluid interfaces in strong
electric/magnetic fields has been seen in various electrospraying
and ferrofluid experiments
\cite{zeleny1917,taylor1964,garton1964,fernandez1992,nagel2000,bacri1982b}.
Building on the work of Taylor \cite{taylor1964}, Li \textit{et al.} and
Ramos \& Castellanos studied 
the electrostatics of an infinite cone with semi-vertical angle
$\theta_0$ formed between two dielectric fluids with permittivity ratio
$\beta$. In spherical coordinates, the electric
stress $\sigma_e\sim r^{2(\nu-1)}$. In an equilibrium cone, this stress
must be balanced by the surface tension, so that $\nu$ must be
$1/2$. According to their analysis \cite{li1994,ramos1994}, there are two
such solutions of $\theta_0$, $\theta_<(\beta)$ and $\theta_>(\beta)$,
which will occur for $\beta > \beta_c=17.59$. The former is
said to be stable, the latter unstable \cite{li1994}. 

In contrast, this letter describes a dynamical fixed point in which a 
cone is formed transiently. At the fixed point, the cone angle is
$\theta_<(\beta)$ so that surface stress and electric stress have the
same scaling in the cone, with the ratio of the two being constant,  but
different from  
unity. The electric stress is the larger of the two, so that the 
total surface stress always acts to elongate the pointed region.

\begin{figure}
\includegraphics[width=\columnwidth]{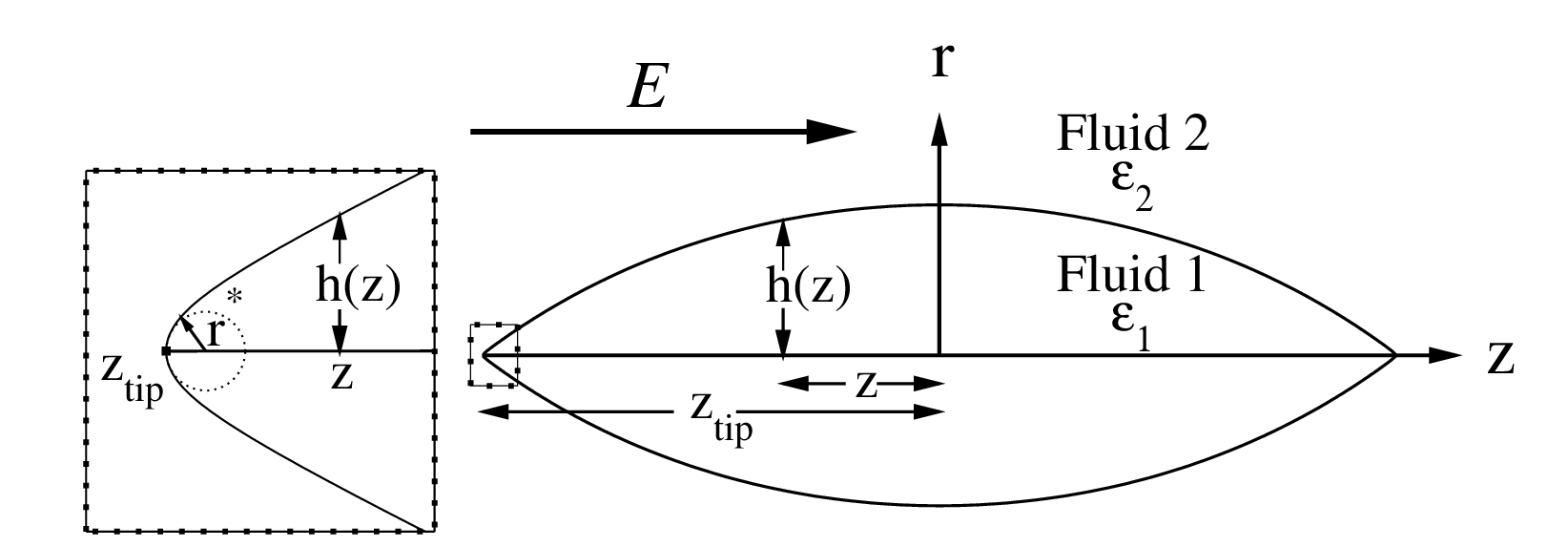}
\caption{A drop of dielectric fluid freely
  suspended in another dielectric fluid in a uniform electric field. } 
\label{fig:schematic}
\end{figure}

We compute the time-dependent deformation of a dielectric drop (fluid
  1) freely suspended in another dielectric fluid (fluid 2) 
in a uniform electric field. Both fluids are incompressible
and have the same viscosity $\eta$. There is surface tension with
coefficient $\gamma$ between the two fluids. The drop is axially
symmetric and has round tips, with its shape represented by the radius
function $h(z)$ in cylindrical coordinates $(r,\ z)$. $r^*$ denotes the
  radius of curvature at the tip. An electric field 
with strength $E$ is applied in the z direction
(Fig. \ref{fig:schematic}). Suppose the initial radius of the drop is $a$.
Respectively we use $a$, $\gamma/\eta$, $\gamma/a$,
  $(\gamma/a\epsilon_o\epsilon_2)^{\frac{1}{2}}$ and
  $(\gamma\epsilon_o/a\epsilon_2)^{\frac{1}{2}}$ to scale length,
  velocity, stress, electric field and surface charge density
  \cite{sherwood1988}. 

Following Sherwood \cite{sherwood1988}, we study a situation in which
Reynolds number is small so that the fluid flows via Stokes equation and
the charge distributions are determined by electrostatics. The
surface charge density $\rho$ can be expressed in the form of a
boundary integral equation 
\begin{eqnarray}
\frac{(\beta+1)}{2(\beta-1)}\rho(x)&=&\int_{L_y}
g(x,y)\rho(y)h(y)(1+h^{'}(y)^2)^{\frac{1}{2}}dy\nonumber\\
& &-\frac{Eh^{'}(x)}{(1+h^{'}(x)^2)^{\frac{1}{2}}},\label{eqn:electric}
\end{eqnarray}
where $\rho(x)$ is the surface charge density at $(h(x),\ x)$, $g$ is a
Green function;  $\beta$
denotes the permittivity ratio $\epsilon_1/\epsilon_2$; $L_y$ is the
range of z axis occupied by the drop \cite{sherwood1988}. From the
surface charge density 
$\rho$, the normal and tangential component of the electric field can be
calculated to obtain the jump in the electric stress across the
interface. Sherwood also uses a boundary integral to determine the
interface velocity 
\begin{equation}
u_i(x)=\frac{1}{8\pi}\int_{L_y}G_{ij}(x,y)f_j(y)h(y)(1+h^{'}(y)^2)^{\frac{1}{2}}dy,\label{eqn:velocity}
\end{equation}
where $i$ and $j$ refer to the $z$ or $r$ component, $f_j(y)$ is the
$j$ component of the total surface stress, and $G$ denotes a Green
function \cite{rallison1978}. The velocity, u, is then used to update
the interface position. In our simulation, we apply a boundary
element method with many details similar to that described by
Sherwood. We distribute mesh points in proportion to the local 
curvature, and use a cubic spline to interpolate the interface
between mesh points, a quartic polynomial to interpolate the surface
charge density. The derived linear algebraic equations are solved
by LU decomposition. A fourth-order Runge-Kutta scheme is applied to update the
interface position. 

There exists a critical electric field $E_c(\beta)$ for $\beta>\beta_c$.
When $E<E_c$, the drop can reach equilibrium with
round tips. We start our simulation from a sphere and apply a
sufficiently small electric field. If the maximum velocity on the
interface decreases to a value below $10^{-4}$ following an exponential
decay, we consider that the drop will reach equilibrium. After equilibrium is
reached by a numerical extrapolation, the field is increased by a small
amount. Through increasing the electric field step by step we find the
critical electric field. 

\begin{figure}
%If possible, increase the width of this graph.
\includegraphics[width=\columnwidth]{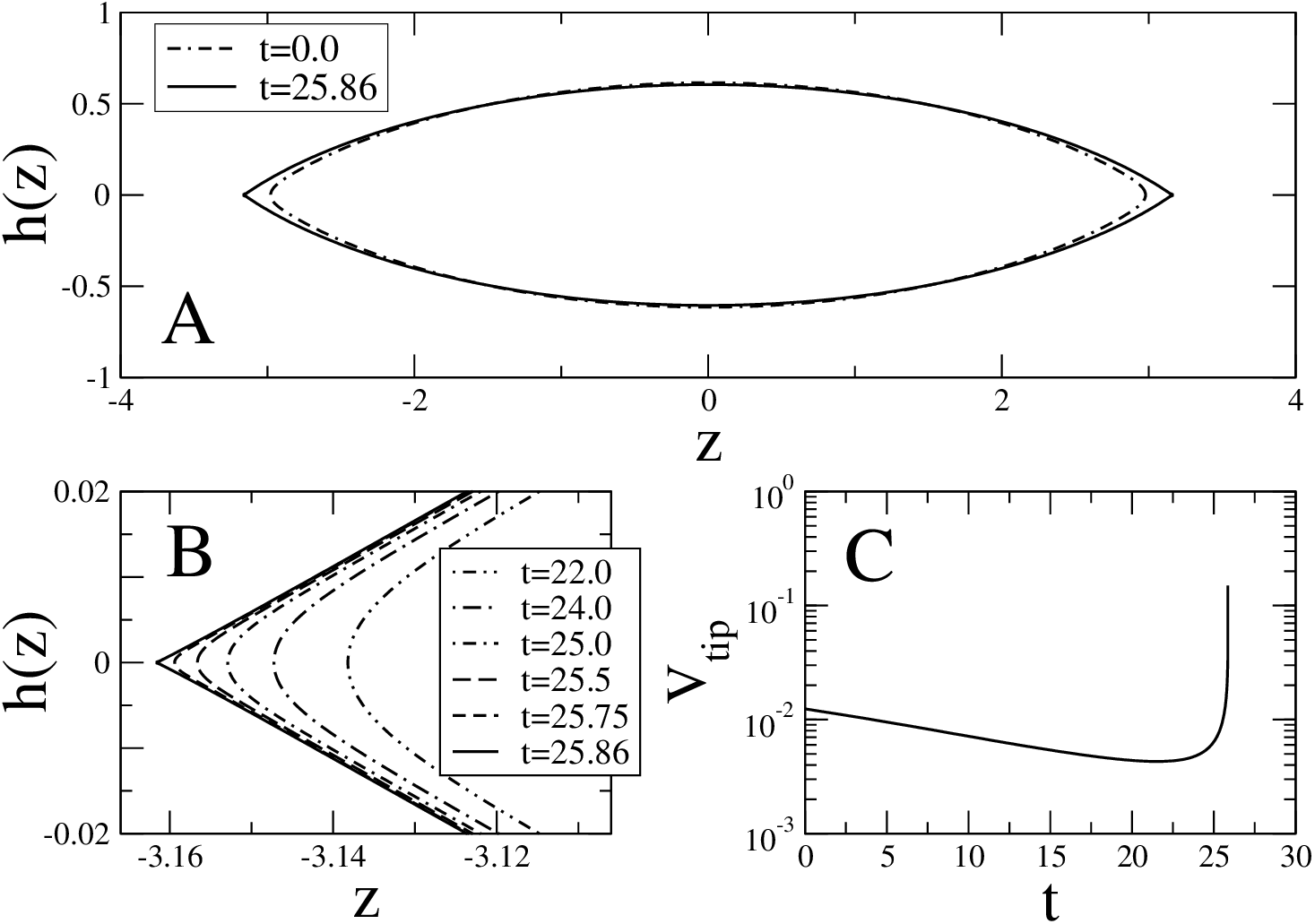}
\caption{Development of a finite singularity at $\beta=18.5 > \beta_c$ and
  $E^2=0.410>E^2_c$. (A) The initial shape (the equilibrium shape
  at $E^2=0.400$), and the final shape calculated ($r^*=10^{-12}$). 
  (B) Formation of conical ends. (C) Diverging velocity at the tip.}
  \label{fig:185conical} 
\end{figure}

\begin{figure}
\includegraphics[width=\columnwidth]{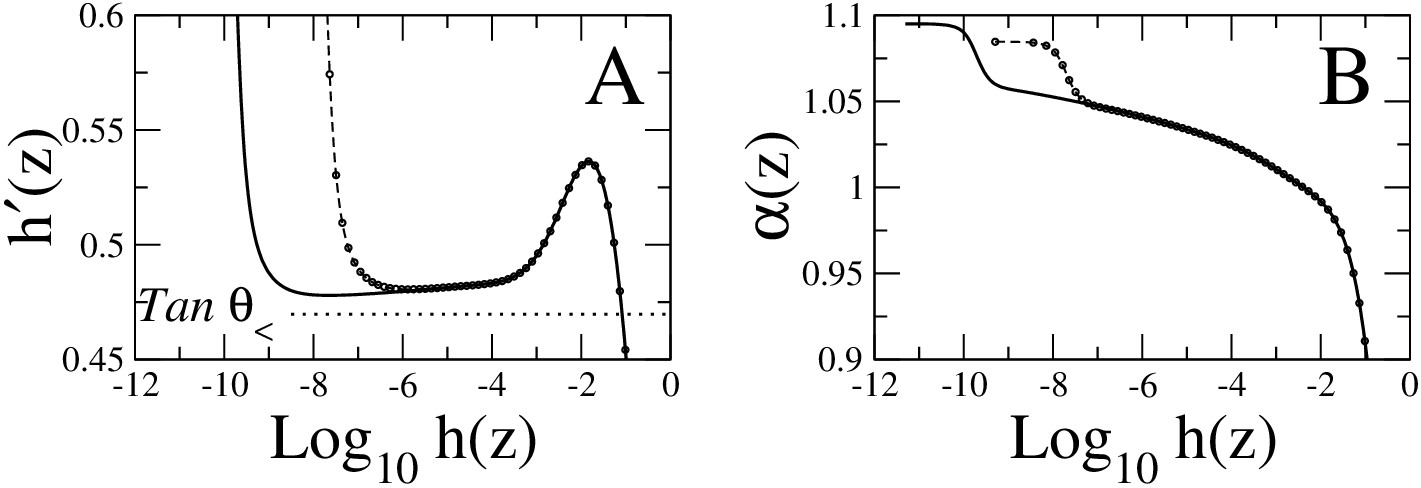}
\caption{The shape and stresses at $r^*=10^{-8}$ (dashed with dots) and
  $r^*=10^{-10}$ (solid) from the simulation in
  Fig. \ref{fig:185conical}. (A) Slopes on the interface. (B) Ratios of
  the electric stress to the surface tension on the interface. $Tan\
  \theta_>=0.689654$ and $Tan \theta_<=0.469704$ at $\beta=18.5$. The later
  curve matches the earlier one on the right hand parts of the plots. } 
\label{fig:185singular}
\end{figure}

When $E>E_c$, a finite time singularity develops. From now on, we use
$\beta=18.5$ as our example, which has $E^2_c=0.4085\pm0.0003$. For
instance, when we choose the initial shape to be the equilibrium shape
at $E^2=0.400 < E^2_c$ and suddenly apply $E^2=0.410 > E^2_c$, the drop
forms conical-like ends. The velocity at the tip dramatically increases
as a critical time is approached (Fig. \ref{fig:185conical}). Here we can at
most obtain about twelve decades of data in one calculation, due to the
increasing number of mesh points required and roundoff error. 

Figure \ref{fig:185singular} shows how the shape and stresses evolve as the
finite time singularity develops. The slope plot suggests we can partition the
interface into three regions, the tip region, the conical-like region
and the macroscopic region. The conical-like region is the intermediate
region with a small variation in the slope. Figure
\ref{fig:185singular}(A) shows as $r^*$ decreases, the macroscopic
region and the established part of the cone region almost remain
intact, while part of the interface which used to 
be in the tip region now grows conical-like. This shows that in the course of
$r^* \to 0$, only the tip region changes in time, while the established
part of the cone region nearly remains independent of time. 
A careful examination of Fig. \ref{fig:185singular}(A) shows the
conical-like region is not exactly a cone, because the slope of the
newly-grown cone changes as $r^*$ decreases. As we shall see
in more detail later, this slope approaches the value set by $\theta_<$. Figure
\ref{fig:185singular}(B) shows the ratio of the electric stress to the
surface tension is larger than one in the tip region and the
conical-like region. The stress ratio also does not change in
the part of interface whose shape remains as the conical-like region
expands. Hence the numerical evidence says that the shape of each part
of the almost conical region and the stress ratio within that part remain
frozen as the tip gets smaller. However, as $r^*$ changes the slope and the
stress ratio of the newly-grown part change too. So Figure \ref{fig:185conical}
and \ref{fig:185singular} may show an approach to a fixed point, but
they do not show a fixed behavior themselves. Because there is a slow
and not-quite uniform convergence to a fixed point, it is hard to
estimate the critical time, $t_c$, from our raw data. For this reason, we shall
henceforth plot our results against tip radius, $r^*$ instead of trying to
use $t_c-t$. 

A scaling study, sometimes called an order of magnitude analysis, enables
us to estimate the sizes of different contributions, it indicates how
the different regions affect one another.
 
These estimates show that except for a uniform advection, the 
stresses in the tip region determines the flow within that region and the 
subsequent shape of the tip. Specifically, the deformation
of the tip region is dominantly caused by the local stress jump. The
axial strain rate defined as $\partial u_z(x)/\partial x$ measures
how fast the interface deforms due to the axial velocity. Using
(\ref{eqn:velocity}), we can express the contribution to $\partial
u_z(x)/\partial x$ from the three regions. Respectively the tip region,
the conical region and the macroscopic region have length scales $r^*$,
$h(z)$ and $1$ ($r^* \ll h(z) \ll 1$). The electric stresses in the
three regions have an order of magnitude 
\begin{equation}
\sigma_e \sim\rho^2\sim\left\{
\begin{array}{ll}
E^2\ r^{*2(\nu-1)} & \mbox{tip region}\\
E^2\ h(z)^{2(\nu-1)} & \mbox{conical-like region,}\\
E^2 & \mbox{macroscopic region}
\end{array}
\right. \label{eqn:stressorder}
\end{equation}
with $0<\nu<1$. A similar result applies to the surface tension
$\sigma_s$, but with $\nu=1/2$. An argument like that of Lister and
Stone shows that the forces in the intermediate and macroscopic region
simply advect the tip region without significant contribution to the
strain rate \cite{lister1998}. 

A followup study \cite{yang2002} will
describe in more detail how the scaling analysis of the electric stress
works. For the present purposes, it suffices to say that the shape in
the tip region mostly determines the electric stress within that region,
except for a coefficient which only depends on the shape in the
other regions and the applied electric field. If we 
change the shape in the other regions, the electric stress in the entire
tip region will be changed by a factor which is independent of the shape
in the tip region. Changing the applied electric field will have the same
effect. So after we reshape the rest part of the interface, we can
restore the electric stress in the tip region by applying a different electric
field of certain strength. 

The scaling study permits us to construct approximate boundary conditions 
which then permits the accurate determination of the subsequent behavior of the
tip. Basically whenever we are 
about to run out of mesh points, we cut off the part of interface far
away from the tip and replace it by a new shape profile which takes fewer
mesh points. Then we restore the electric stress in the entire tip region,
which we can accomplish by adjusting the applied electric field to
restore the electric stress at the tip to its value prior to the
truncation. The tip regions of the prescribed 
new drop and the original drop will subsequently evolve in 
the same way, because the deformation of the tip region is primarily
driven by the local stress jump. We define the rescaled axial distance
$\xi$ and radius function $H(\xi)$ as 
\begin{equation}
\xi=(z-z_{tip})/r^*,\qquad H(\xi)=h(z)/r^*. \label{eqn:rescale}
\end{equation}
We at least keep the part of interface with $\xi\le 10^{4}$ and
typically match a spherical band to the center region, requiring the
slope to be continuous at the truncation points. The center of the
spherical band, which locates on the z axis, coincides with the center
of the prescribed new drop. The error will be smaller if the truncation
point is farther away from the tip. The ``truncate and prescribe'' idea
was invented by Zhang and Lister \cite{zhang1999}.

\begin{figure}
\includegraphics[width=2.75in]{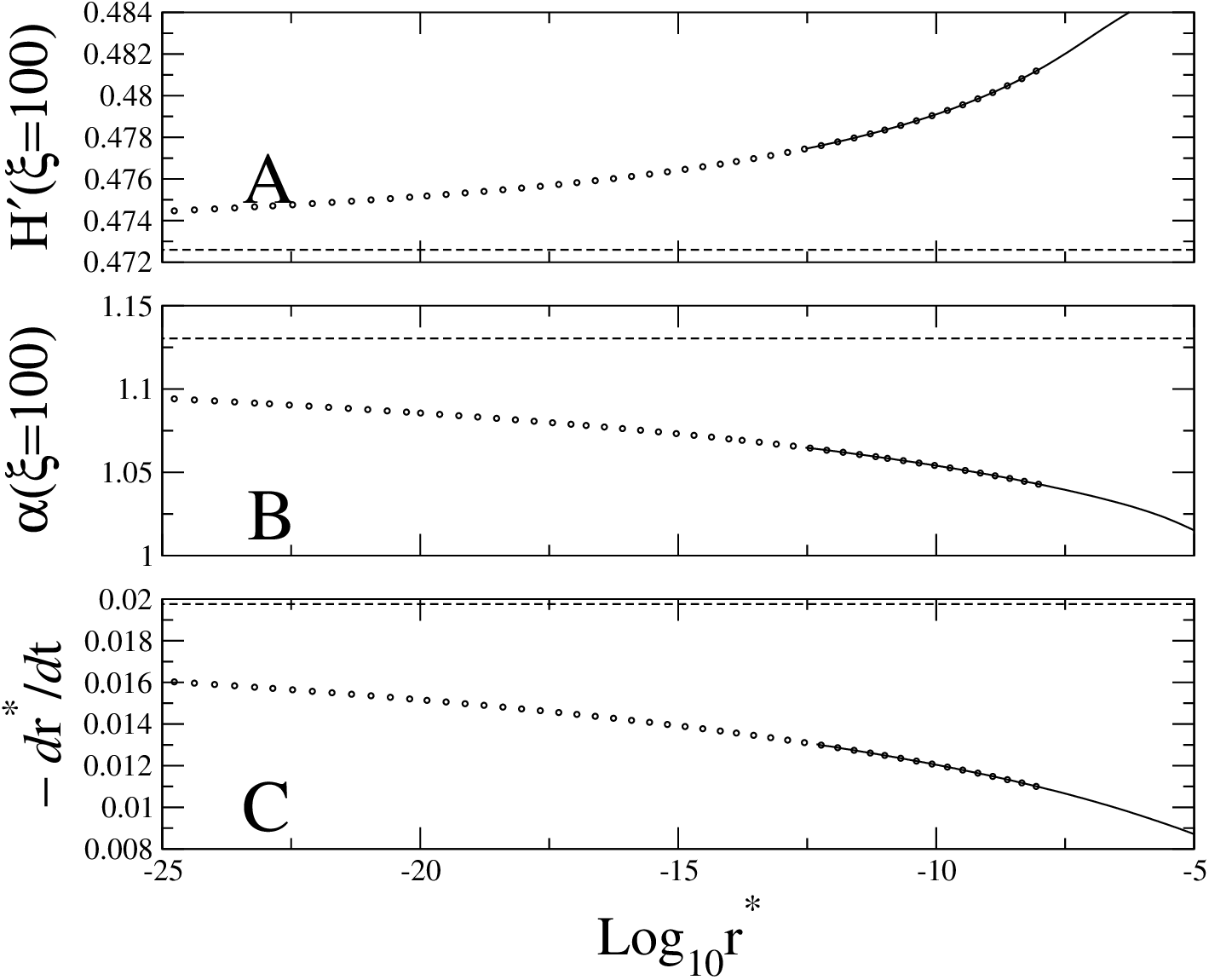}
\caption{Overlap between the results calculated with the exact boundary
  condition (solid curves) and the approximate boundary condition
  (dotted curves). The initial condition is given in
  Fig. \ref{fig:185conical}. Evolution of the tip region is 
  calculated for totally $80$ decades of $r^*$, only part of which are
  shown here. Out of the $80$ decades of data, the curves at
  small $r^*$ converges as power laws in $r^*$. The fit shows as $r^*
  \to 0$: (A) The slope at $\xi=100$ converges to $0.47260\pm 0.00003$, (B) the
  stress ratio at $\xi=100$ converges to $1.1302\pm 0.0004$, and (C) $dr^*/dt$
  converges to $-0.01977 \pm 0.00003$.}  

\label{fig:185sto80}
\end{figure}

\begin{figure}
\includegraphics[width=2.75in] {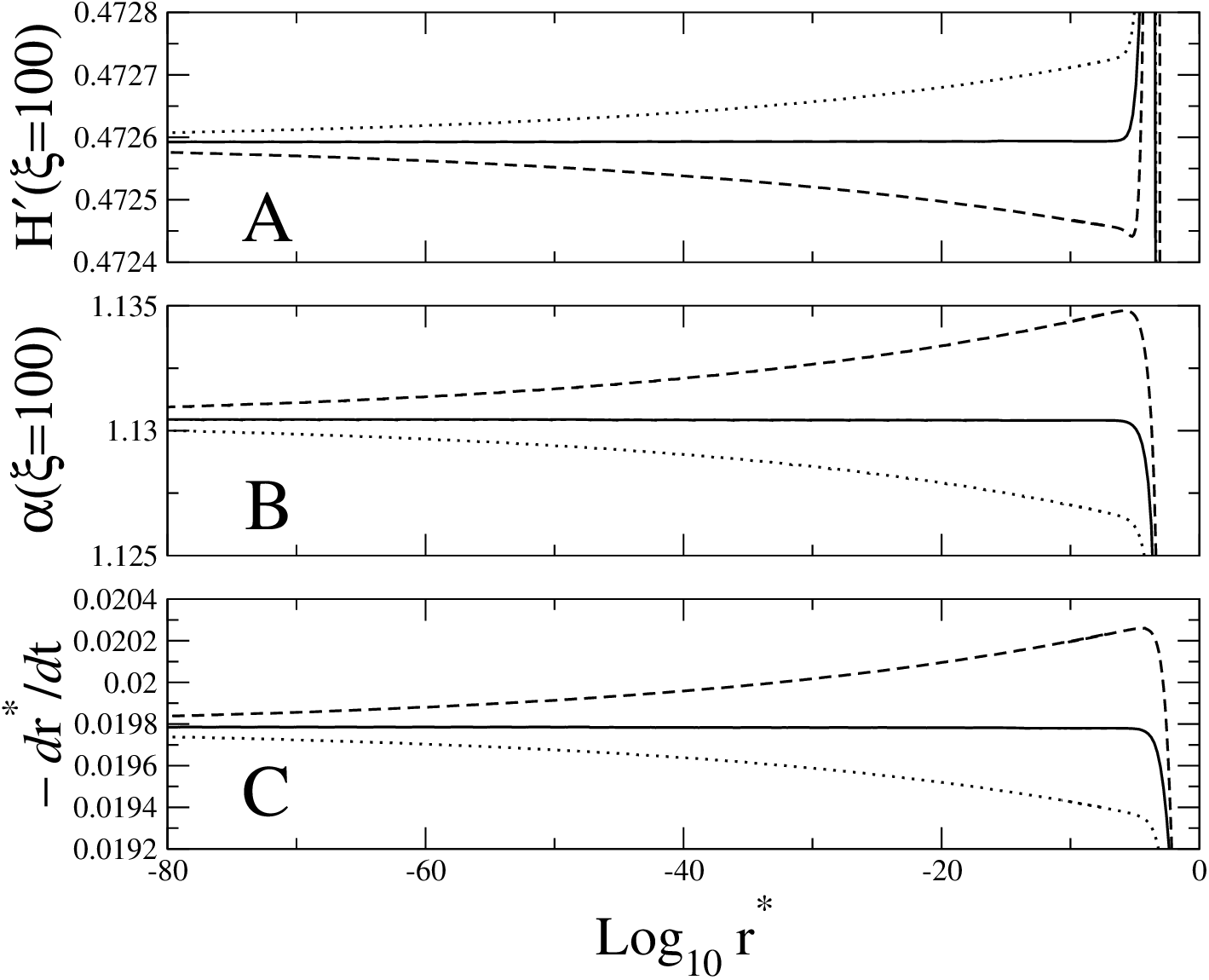} 
\caption{Convergence to a fixed point at $\beta=18.5$. The initial
  shape and the applied field of simulation (i) in solid lines: a sphere,
  $E^2=0.4748$; simulation (ii) in dotted lines: a sphere, $E^2=0.473$;
  simulation (iii) in dashed lines: the equilibrium shape at
  $E^2=0.408$, currently $E^2=0.478$. The overlap between the
  approximate boundary condition results and the exact boundary
  condition results is similarly checked like in
  Fig. \ref{fig:185sto80}. As $r^* \to 0$, (A) 
  $H'(\xi=100)\to 0.472595 \pm 0.000005$; (B) $\alpha(\xi=100) \to
  1.13040 \pm 0.00004$; (C) $dr^*/dt \to -0.019780\pm 0.000003$.}  
\label{fig:185tofix}
\end{figure}

Using the same initial condition as in Fig. \ref{fig:185conical}, we
calculate the evolution of the tip region for eighty decades of $r^*$ with 
approximate boundary conditions. We truncate the drop  
for twenty-four times, starting at $r^*=10^{-8}$. Figure
\ref{fig:185sto80} shows that the approximate boundary condition produces
the same result as the exact boundary condition without truncation at
$10^{-12}\le r^*\le 10^{-8}$. The point with $\xi=100$ is pretty close
to the cone region, so the slope and the stress ratio there can
respectively reflect the angle and the stress ratio of the newly-grown
cone. Out of the eighty decades of data, the curves at small $r^*$ can be
adequately fitted as $c+b(r^{*})^p$ with $p > 0$. Thus $H^{'}(\xi=100)$,
$\alpha(\xi=100)$ and $dr^*/dt$ each approach limiting values as $r^* \to
0$. Just below we shall show those limits are the same for different
initial conditions.  

Further simulations show there exists a fixed point behavior: the
stress ratio of the newly-grown cone converges to a fixed value
larger than unity. At $\beta=18.5$, $\alpha(\xi=100)$ will converge to
$1.130$ as $r^*\to 0$, if it is close to $1$ when the drop starts to
develop conical ends, regardless of the initial shapes. For example in
Fig. \ref{fig:185tofix}, simulation (ii) in solid lines and simulation
(iii) in dashed lines have different initial conditions. The dotted
lines and dashed lines can all be excellently fitted by $c+b(r^{*})^p$. As
$r^*\to 0$, the two simulations give the same limits
$H^{'}(\xi=100)_c=0.4726$, 
$\alpha(\xi=100)_c=1.130$ and $(dr^{*}/dt)_{c}=-0.0198$. As you may
notice, we have obtained the same limits in the simulation in
Fig. \ref{fig:185sto80}. The fitted values of the 
power are very close to each other in all the simulations, and we
get $p=0.013 \pm 0.001$. In simulation (i), we 
purposely choose a particular initial condition to let
$\alpha(\xi=100)$ equal to $\alpha(\xi=100)_c$ at an early stage in the
cone formation. We see that $\alpha(\xi=100)$ stably stays at
$\alpha(\xi=100)_c$ for many decades of $r^*$ with $H^{'}(\xi=100)$ and
$dr^{*}/dt$ equal to the limits obtained from the curve fitting. We have
concrete numerical evidence that $\alpha(\xi=100)_c=1.130$ is a stable
fixed point at $\beta=18.5$. This fixed point is finally approached in a
power law of $r^*$. And simulation (i) gives the solution at this fixed point. 

\begin{figure}
\includegraphics[width=\columnwidth]{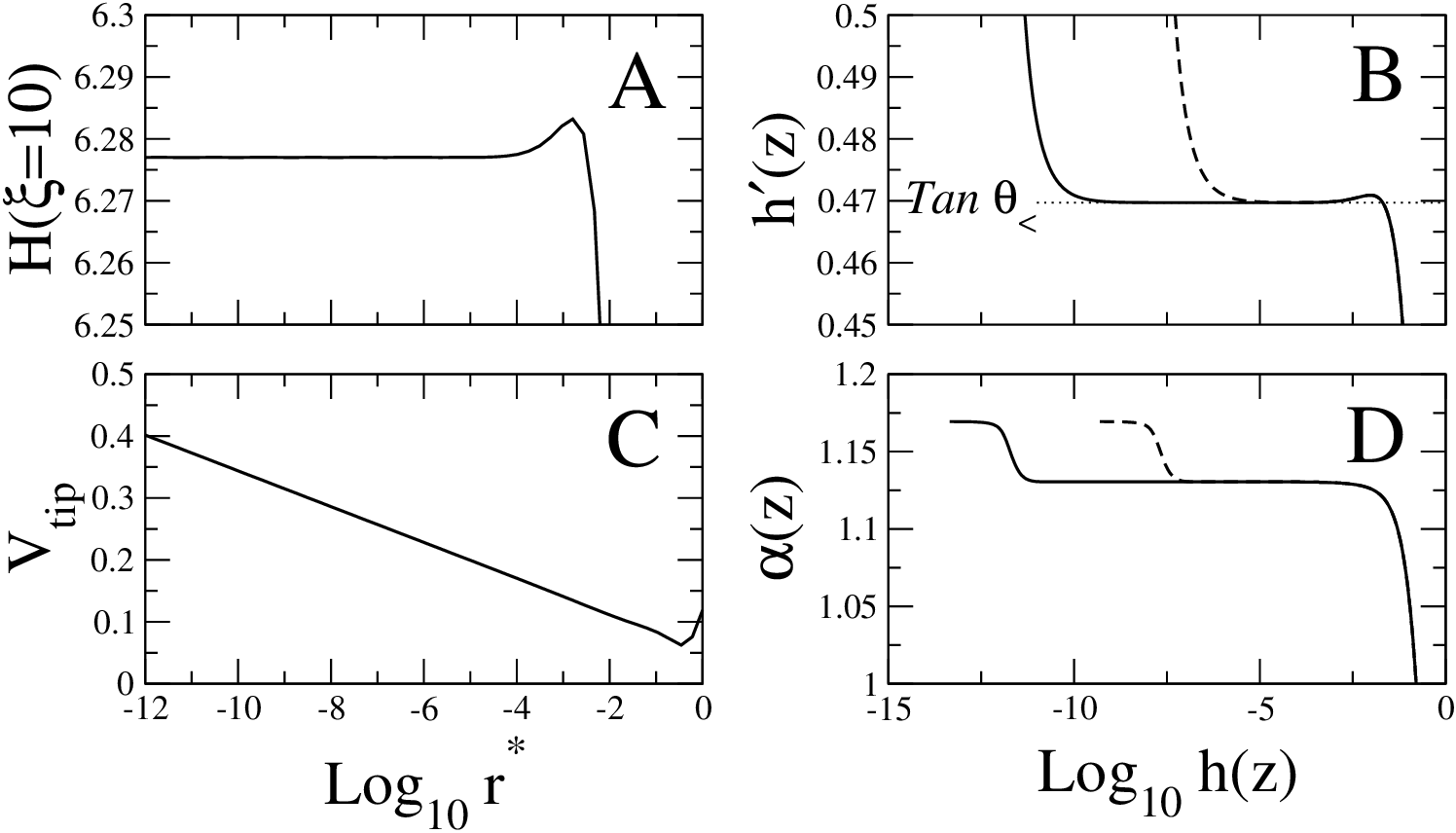}
\caption{The self-similar solution at the fixed point $\alpha(\xi=100)_c=1.130$
  at $\beta=18.5$. The first $12$ decades of data computed with the
  exact boundary condition in simulation (i) reveal: (A) $H(\xi)$ is a constant
  for small $r^*$, for example at $\xi=10$. (B)
  The intermediate region has a constant slope $0.46970$
  nearly equal to $Tan\ \theta_<(18.5)=0.469704$. (C) The velocity at
  the tip increases 
  logarithmically. (D) The stress ratio in the intermediate region is a 
  constant $1.130 > 1$. In (B)(D), dashed curves: $r^*=10^{-8}$, solid
  curves: $r^*=10^{-12}$.} 
\label{fig:185fix}
\end{figure}

At this fixed point, the tip region is self-similar and the
intermediate region is a cone with the cone angle $\theta_<(\beta)$. The 
first twelve decades of data calculated with the exact boundary
condition in simulation (i) reveal the solution at this fixed point. We
find the following properties: (a) The 
shape profiles of the tip region are self-similar after we rescale them 
by $r^{*}$, for example $H(\xi)$ at $\xi=10$ is a constant for small $r^*$
[Fig. \ref{fig:185fix}(A)]. (b) The ever expanding intermediate region 
has a constant slope equal to $Tan \ \theta_{<}(\beta)$
[Fig. \ref{fig:185fix}(B)]. (c) The velocity at 
the tip increases logarithmically [Fig. \ref{fig:185fix}(C)]. (d) The 
stress ratio in the intermediate region is a constant substantially
larger than one, which we call it $\alpha_c$
[Fig. \ref{fig:185fix}(D)]. The values of $\alpha_c$ and
$\alpha(\xi=100)_c$ are very close to each other. (e) $dr^{*}/dt$ is a 
constant [Fig. \ref{fig:185tofix}(C)], which indicates that $r^*$ scales
like $t_c-t$. This self-similar solution has some qualitative similarity
with a scaling solution found by Lister and Stone \cite{lister1998}.

We find similar fixed points at other values of $\beta>\beta_c$ such as
$19.0$. To summarize, our numerical evidence shows a cone with the
smaller cone angle $\theta_<(\beta)$ can be formed transiently in a
non-equilibrium case where the electric stress is not balanced by the
surface tension. The angle is approached as the stress ratio in the
newly-grown cone region converges to a fixed value $\alpha_c(\beta)>1$. The
dynamical solution at this fixed point is self-similar. 

This project would be simply impossible without Leo P. Kadanoff's
guidance and support. I am very grateful to H. A. Stone and W. W. Zhang for
their codes on viscous pinchoff. I also want to thank W. W. Zhang for
helpful discussions. This research was supported in part by the DOE
ASCI-FLASH program, the NSF grant DMR-0094569 to L.
P. Kadanoff, and the MRSEC program of the National Science Foundation
under Award No. DMR-9808595.
%\bibliography{newfixed}

\end{document}